\documentclass[12pt]{iopart}
\usepackage{iopams}
\usepackage{epsf}
\usepackage{epsfig}
\newlength{\myspace}
\newcommand{\fr}{{^F\hspace{-.02in}R}}
\begin{document}
\title{Tidal Dynamics in Kerr Spacetime}

\author{C. Chicone$^1$ and B. Mashhoon$^2$}

\address{$^1$ Department of Mathematics, University of Missouri-Columbia, Columbia, MO 65211, USA}
\address{$^2$ Department of Physics and Astronomy, University of Missouri-Columbia, Columbia, MO 65211, USA}

\begin{abstract}
The motion of free nearby test particles relative to a stable equatorial circular geodesic orbit about a Kerr source is investigated. It is shown that the nonlinear generalized Jacobi equation can be transformed in this case to an autonomous form. Tidal dynamics beyond the critical speed $c/ \sqrt{2}$ is studied. We show, in particular, that a free test particle vertically launched from the circular orbit parallel or antiparallel to the Kerr rotation axis is tidally accelerated if its initial relative speed exceeds $c/\sqrt{2}$.  Possible applications of our results to high-energy astrophysics are briefly mentioned.
\end{abstract}
\pacs{04.20.Cv}
\maketitle

\section{Introduction}

The purpose of this paper is to study some aspects of the relative motion of free test particles in the gravitational field of a rotating astronomical source. Imagine, for the sake of concreteness, a free mass $m$ in orbit about the central source of mass $M\gg m$ and angular momentum $J$, where $m$, $M$ and $J$ are constants. Let $\tau$ be the proper time along the orbit of mass $m$ and $\lambda^\mu_{\hspace{\myspace}(\alpha)}$ be a local orthonormal tetrad system that is parallel transported
along this path. Here $\lambda^\mu_{\hspace{\myspace}(0)}=dx^\mu/d\tau$ is the four-velocity vector of $m$ and $\lambda^\mu_{\hspace{\myspace}(i)}$, $i=1,2,3$,  are unit spacelike gyro directions that determine the local spatial frame of $m$. We employ units such that $ c = G = 1$ throughout; moreover, the signature of
the metric is $+2$ in our convention.  Suppose that at $\tau =0$ a probe is launched  from $m$ and for some time $\tau >0$ the motion of the probe is closely monitored by observers comoving with $m$. To study relative motion in an invariant manner within the context of general relativity, the quasi-inertial Fermi normal coordinate system \cite{1} is indispensable as it most closely corresponds to actual observations. Let $X^\alpha=( T, \mathbf{X} )$ be the Fermi coordinates of the probe at an event $P$, then there
exists a unique spacelike geodesic of length $\ell$ that connects $P$
orthogonally to the path of $m$  at an event $O$ with proper time $\tau$ such that $T=\tau$ and $X_i=\ell n_\mu \lambda^\mu_{\hspace{\myspace}(i)}$, where $n^\mu =(dx^\mu/d\ell)_O$ is the unit vector at $O$ that is tangent to the
spacelike geodesic at $O$ and orthogonal to the worldline of $m$. The metric in
Fermi coordinates is given by
 \begin{eqnarray}
\label{eq12a} g_{00}&=&-1-\fr_{0i0j} X^iX^j+\cdots,\\
\label{eq13a} g_{0i}&=&-\frac{2}{3}\, {\fr_{0jik}}X^jX^k+\cdots,\\
\label{eq14a} g_{ij}&=&\delta_{ij}-\frac{1}{3}\, {\fr_{ikjl}}X^kX^l
                       +\cdots
\end{eqnarray}
to second order in the distance away from $m$ that permanently lies
at the spatial origin of the Fermi coordinates by construction and has Fermi
coordinates $(\tau,\mathbf{0})$.
Here ${\fr_{\alpha \beta \gamma \delta}}$ is defined by
\begin{equation}
\label{eq5} {\fr_{\alpha \beta \gamma \delta}} =R_{\mu\nu \rho \sigma} \lambda^\mu_{\hspace{\myspace} (\alpha)} \lambda^\nu_{\hspace{\myspace}(\beta)} \lambda^\rho_{\hspace{\myspace}(\gamma )} \lambda^\sigma_{\hspace{\myspace} (\delta)},\end{equation}
which is the projection of the Riemann curvature tensor on the tetrad frame $\lambda^\mu _{\hspace{\myspace} (\alpha)}$ along the reference worldline.
The theoretical equation of motion of the probe relative to mass $m$ can be described by the reduced geodesic equation~\cite{2}
\begin{equation}
\label{eq1} \frac{d^2X^i}{dT^2} + (\Gamma^i_{\alpha \beta} -\Gamma ^0_{\alpha \beta} V^i) \frac{dX^\alpha}{dT} \frac{dX^\beta}{dT}=0, 
\end{equation}
where $\mathbf{V}=d\mathbf{X}/dT$ is the Fermi velocity of the probe relative to $m$. The four-velocity of the probe is given by $U^\mu=\Gamma (1,\mathbf{V})$, where $\Gamma$ can be determined from
\begin{equation}
\label{eq2} -\frac{1}{\Gamma^2}=g_{00}+2g_{0i} V^i+g_{ij} V^iV^j<0, 
\end{equation}
since $U^\mu$ is a timelike unit vector. The Fermi coordinates are geodesic coordinates based on an orthonormal tetrad frame $\lambda^\mu_{\hspace{\myspace} (\alpha )} (\tau )$ that is parallel propagated along the orbit of $m$. They are admissible within a cylindrical region of radius $\mathcal{R}_a$ along the worldline of $m$; that is, $|\mathbf{X}|<\mathcal{R}_a$, where $\mathcal{R}_a(T)$ is a certain radius of curvature of spacetime. The nature of $\mathcal{R}_a(T)$ has been discussed in detail in our recent paper on explicit Fermi coordinates~\cite{3}.

For the sake of simplicity, we limit our considerations to $|\mathbf{X}|\ll \mathcal{R}_a$. A general feature of the reduced geodesic~\eref{eq1} is that it contains a critical speed given by $1/\sqrt{2}$, especially in the case of one-dimensional motion. The notion of critical speed in gravitational motion has been recently reviewed in ref.~\cite{newbm}. We are particularly interested in the role that the critical speed $V_c =1/\sqrt{2}$ plays in the motion of the probe; therefore the probe can be launched from $m$ at any initial speed $V_0<1$, but---for simplicity---we concentrate on its motion only when it is relatively close to the reference particle $m$. \Eref{eq1} to first order in $|\mathbf{X}|$ is the generalized Jacobi equation~\cite{2}
\begin{eqnarray}
\fl \frac{d^2X^i}{dT^2} +{\fr_{0i0j}}X^j+2  {\fr_{ikj0}} V^kX^j\nonumber\\
+\frac{2}{3} (3 {\fr_{0kj0} }V^iX^k+  {\fr_{ikj\ell}}V^kV^\ell +  {\fr_{0kj\ell} }V^i V^kV^\ell )X^j=0,\label{eq3}
\end{eqnarray}
which describes the timelike motion of the probe when
\begin{eqnarray} 
\fl \frac{1}{\Gamma^2} = 1-V^2+\; {\fr_{0i0j}} X^i X^j+\frac{4}{3} {\fr_{0jik}} X^j V^i X^k
+\frac{1}{3}  {\fr_{ikj\ell}} V^iX^kV^jX^\ell >0.\label{eq4}
\end{eqnarray}
 For $|\mathbf{V}|\ll 1$, we can drop all velocity-dependent terms in equations~\eref{eq3} and \eref{eq4}; then, the inequality in~\eref{eq4} is satisfied and \eref{eq3} reduces to the Jacobi equation
\begin{equation}\label{eq6} \frac{d^2X^i}{dT^2} +\; {^F R_{0i0j}} X^j=0.\end{equation}

Suppose that the worldline of the probe can be determined in the Fermi
coordinate system on the basis of equations~\eref{eq1} and~\eref{eq2}; then, it is in
principle possible to find the representation of this worldline in the
background coordinate system using the explicit coordinate transformation
between the two systems of coordinates. In practice, however, it is not
possible in general to find explicit expressions for Fermi coordinates in
terms of the background coordinates and therefore one has to resort to
approximation schemes; see ref.~\cite{3}  for a general discussion of this
problem. Nevertheless, it is important to recognize that once $\mathbf{X}(T)$ is determined from the generalized Jacobi equation (or the Jacobi equation), the worldline of the probe is completely characterized. That is, if in the background coordinate system the worldline of $m$ is given by $x^\mu(\tau)$, then the worldline of the probe is $x_p^\mu(\tau)$, where

\begin{eqnarray}\label{eq:10}
x_p^\mu(\tau)=x^\mu(\tau)+X^i(\tau) \lambda^\mu_{\hspace{\myspace}(i)}(\tau)
\end{eqnarray}
and we have replaced the Fermi temporal coordinate $T$ by $\tau$ in the argument of $\mathbf X$. 
This relation follows from the definition of Fermi coordinates and the fact that $x^\mu_p(\tau)-x^\mu(\tau)\approx \ell n^\mu$ for $\ell\ll \mathcal{R}_a$,  since the (generalized) Jacobi equation is linear in the distance away from the reference trajectory.
Alternatively, $x_p^\mu(\tau)$ may be viewed as a perturbed geodesic orbit. Depending upon whether $\mathbf X$ is determined from the Jacobi or the generalized Jacobi equation, the results of this approach may be interpreted respectively in terms of the linear or nonlinear perturbations of geodesic orbits in the background gravitational field.

Regarding the source of the gravitational field and the reference trajectory, we assume that $m$ is on a stable equatorial circular geodesic orbit about a Kerr source.
This choice for the orbit to some extent complements our previous work on
the motion of astrophysical jets~\cite{4,5}, where the reference trajectory is a
radial geodesic along the rotation axis of a Kerr source. The geodesics of Kerr spacetime have been discussed by a number of authors; see, for instance, ref.~\cite{6} and references therein. The characteristics of the source and the reference trajectory will be described in detail in the next section. Various aspects of tidal dynamics in black-hole
spacetimes have been discussed in a number of papers; see, for instance~\cite{7,8,9,10,11}  and the references cited therein. Furthermore, we treat free-particle orbits for the sake of simplicity; a corresponding discussion of black-hole accretion disks~\cite{15} is beyond the scope of this investigation. Section~\ref{s3} is devoted to the solution of the Jacobi equation. We show in section~\ref{s4} that a rotation introduced in section~\ref{s3} leads to a complete transformation of the generalized Jacobi equation to an autonomous form. This equation is then analyzed in detail. Finally, section~\ref{s5} contains a brief discussion of our results.

\section{Reference Worldline\label{s2}}

The spacetime region of interest is the exterior Kerr domain with the metric
\begin{eqnarray}
-ds^2 &=-dt^2+\frac{\Sigma}{\Delta} dr^2 +\Sigma d\theta^2+(r^2+a^2) \sin^2 \theta d\phi^2\nonumber\\
&\quad + \frac{2Mr}{\Sigma} (dt-a\sin^2 \theta d\phi )^2,\label{eq7}
\end{eqnarray}
where $a=J/M$ is the specific angular momentum of the source. Here $(t,r,\theta ,\phi)$ are the standard Boyer-Lindquist coordinates and
\begin{equation} \label{eq8} \Sigma =r^2 +a^2\cos ^2\theta,\quad \Delta =r^2-2Mr+a^2.\end{equation}
The reference trajectory is assumed to be a stable circular orbit of fixed radius $r$ in the equatorial plane $\theta=\pi /2$; in terms of $r$, such orbits exist from infinity all the way down to the last stable circular orbits given by
\begin{equation}
\label{eq9} 1-\frac{6M}{r}\pm 8a\sqrt{\frac{M}{r^3}} -3\frac{a^2}{r^2}=0.
\end{equation}
Throughout this paper, the upper (lower) sign refers to orbits where $m$ rotates in the same (opposite) sense as the source. For $r$ less than the solution of \eref{eq9}, there are unstable circular orbits that end at the null circular orbits given by
\begin{equation}
\label{eq10} 1-\frac{3M}{r}\pm 2a\sqrt{\frac{M}{r^3}}=0.
\end{equation}

It is useful to define the Keplerian frequency for the orbits under consideration here as
\begin{equation}
\label{eq11} \omega _K=\pm \sqrt{\frac{M}{r^3}}.
\end{equation}
Let us note that in the orbital equations \eref{eq9} and \eref{eq10}, a prograde orbit becomes retrograde and vice versa when $a\to -a$; this circumstance explains why the combination $a\omega_k$ usually appears in orbital equations. We assume that $a\geq 0$ throughout; therefore, the sign of $\omega_K$ indicates the sense of the orbit. It follows from the geodesic equation that along the reference worldline
\begin{equation}
\label{eq12} t=\frac{1+a\omega_K}{N} \tau ,\quad \phi =\frac{\omega_K}{N}\tau,
\end{equation}
where
\begin{equation}
\label{eq13} N=\sqrt{1-\frac{3M}{r}+2a\omega_K}
\end{equation}
is such that the null circular orbits are solutions of $N=0$ and we have assumed in \eref{eq12} that $t=\tau =0$ at $\phi =0$. The constants of the motion for the reference worldline are the specific energy $E$ and orbital angular momentum $L$, associated respectively with the timelike and azimuthal Killing vectors $\partial_t$ and $\partial_\phi$, and are given by
\begin{equation}
\label{eq14} E=\frac{1}{N} \left( 1-\frac{2M}{r} +a\omega_K\right),\quad L=\frac{r^2\omega_K}{N}  \left( 1-2a\omega_K+\frac{a^2}{r^2}\right).
\end{equation}

To determine the orthonormal tetrad $\lambda^\mu_{\hspace{\myspace} (\alpha)}$, let us first consider the tetrad $\Lambda^\mu_{\hspace{\myspace}(\alpha)} $ carried by the fundamental static observers in the exterior Kerr spacetime. In the equatorial plane and in terms of $(t, r,\theta,\phi)$, the natural orthonormal tetrad of these observers along the coordinate directions is given by
\begin{eqnarray} 
\label{eq15}\Lambda^\mu _{\hspace{\myspace} (0)} = \left( \frac{1}{A} ,0,0,0 \right) ,\qquad \Lambda ^\mu _{\hspace{\myspace} (1)}= \left( 0,\frac{\sqrt{\Delta}}{r}, 0,0\right),\\
\label{eq16}\Lambda^\mu_{\hspace{\myspace}(2)} =\left( 0,0,\frac{1}{r},0\right), \qquad \Lambda^\mu _{\hspace{\myspace}(3)}= \left( -\frac{2Ma}{rA\sqrt{\Delta}} ,0,0,\frac{A}{\sqrt{\Delta}} \right),
\end{eqnarray}
where
\begin{equation}
\label{eq17}A=\sqrt{1-\frac{2M}{r}}.
\end{equation}
Next, we subject this tetrad to a Lorentz boost with speed $\tilde{\beta}$, $\Lambda^\mu _{\hspace{\myspace}(\alpha)} \to \tilde{\Lambda}^\mu _{\hspace{\myspace}(\alpha)}$, such that $\tilde{\Lambda}^\mu _{\hspace{\myspace}(0)} =\lambda^\mu _{\hspace{\myspace}(0)} $ is the unit vector tangent to the worldline of the reference particle $m$. The new orthonormal tetrad along the orbit of $m$ is
\begin{eqnarray}
\label{eq18} \tilde{\Lambda}^\mu _{\hspace{\myspace} (0)} =\tilde{\gamma } [\Lambda^\mu_{\hspace{\myspace} (0)}+\tilde{\beta}\Lambda^\mu _{\hspace{\myspace}(3)}], \quad  \tilde{\Lambda}^\mu _{(1)} =\Lambda^\mu _{\hspace{\myspace} (1)},\\
\label{eq19}\tilde{\Lambda}^\mu_{\hspace{\myspace}(2)} = \Lambda^\mu _{\hspace{\myspace}(2)}, \quad \tilde{\Lambda}^\mu_{\hspace{\myspace} (3)} = \tilde{\gamma} [\Lambda^\mu_{\hspace{\myspace} (3)} +\tilde{\beta} \Lambda^\mu_{\hspace{\myspace}(0)} ],
\end{eqnarray}
where $\tilde{\gamma}$ is the Lorentz factor corresponding to $\tilde{\beta}$. We find from $\tilde{\Lambda}^\mu_{\hspace{\myspace}(0)} =\lambda^\mu_{\hspace{\myspace}(0)}$ that the Lorentz pair $(\tilde{\beta} ,\tilde{\gamma})$ is given by
\begin{equation}
\label{eq20} \tilde{\beta} =\frac{\sqrt{\Delta}\,\omega_K}{EN},\quad \tilde{\gamma} =\frac{E}{A}.
\end{equation}
\begin{figure}
\centerline{\psfig{file=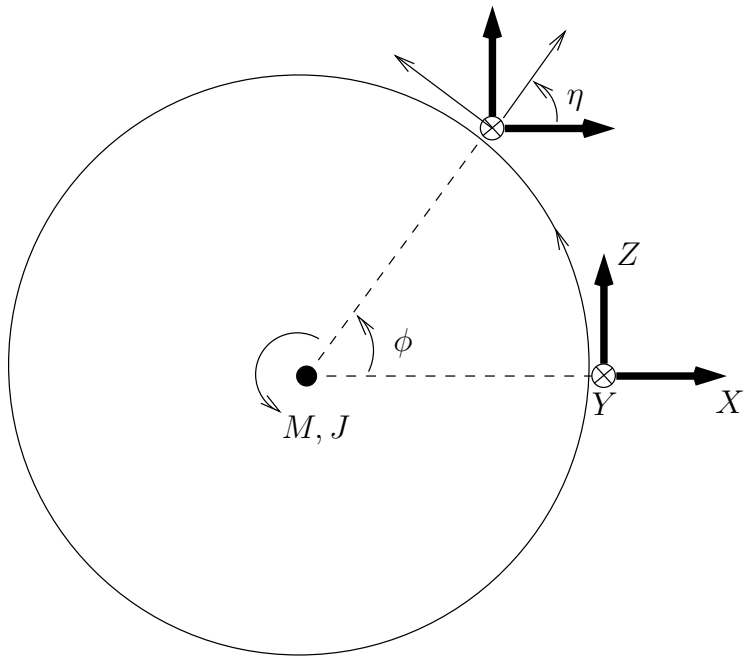, width=25pc}}
\caption{Schematic drawing that depicts the parallel transport of the spatial triad along a prograde circular orbit.
\label{fig:1}}
\end{figure}
Clearly, $\tilde\beta$ is positive (negative) for prograde (retrograde) orbits; moreover, $\tilde\gamma$ diverges at the null orbits ($N=0$). The spatial triad of $\tilde{\Lambda}^\mu_{\hspace{\myspace}(\alpha)}$ is along the spherical polar coordinate directions, which are the radial, normal and tangential directions with respect to the orbit, and therefore needs to be rotated back such that the resulting tetrad would then be parallel propagated along the orbit as illustrated in Figure~\ref{fig:1}. Thus
\begin{eqnarray}
\label{eq21} \lambda^\mu _{\hspace{\myspace}(0)} =\tilde{\Lambda}^\mu _{\hspace{\myspace} (0)},\qquad \lambda^\mu _{\hspace{\myspace} (1)}=\tilde{\Lambda}^\mu_{\hspace{\myspace} (1)} \cos \eta -\tilde{\Lambda}^\mu _{\hspace{\myspace} (3)}\sin \eta ,\\
\label{eq22} \lambda^\mu _{\hspace{\myspace} (2)} =\tilde{\Lambda}^\mu _{\hspace{\myspace} (2)},\qquad \lambda^\mu _{\hspace{\myspace} (3)} =\tilde{\Lambda}^\mu _{\hspace{\myspace} (1)} \sin \eta +\tilde{\Lambda}^\mu _{\hspace{\myspace} (3)} \cos \eta .
\end{eqnarray}
It follows from the vanishing of the covariant derivative of $\lambda^\mu_{\hspace{\myspace} (i)}$ along the orbit that
\begin{equation}
\label{eq23} \eta =\omega_K\tau ,
\end{equation}
where we have chosen the integration constant such that $\eta =0$ at $\tau =0$. 
Note that $\phi-\eta=(N^{-1}-1)\eta$, which, to first order in $M/r \ll 1$ and $a/r \ll 1$ can be written as 
\begin{equation}\label{eq24}
\phi-\eta\approx \big(\frac{3}{2}\frac{M}{r} \omega_K-\frac{J}{r^3}\big)\tau;
\end{equation}
that is, the difference between these angles is due to a combination of
geodetic and gravitomagnetic precessions. The final result for the tetrad frame in $(t,r,\theta,\phi)$ coordinates is then 
\begin{eqnarray}
\label{eq24a} \lambda^\mu_{\hspace{\myspace} (0)} =\frac{1}{N} (1+a\omega_K,0,0,\omega_K),\\
\label{eq25} \lambda^\mu_{\hspace{\myspace}(1)} =\frac{1}{\sqrt{\Delta}} \left( -L\sin \eta ,\frac{\Delta }{r} \cos \eta ,0,-E\sin \eta \right),\\
\label{eq26} \lambda^\mu_{\hspace{\myspace}(2)} =\left( 0,0,\frac{1}{r} ,0\right),\\
\label{eq27} \lambda^\mu_{\hspace{\myspace}(3)} =\frac{1}{\sqrt{\Delta}} \left( L\cos \eta ,\frac{\Delta}{r}\sin \eta ,0,E\cos \eta \right).
\end{eqnarray}

Finally, we need to compute the projection of the Riemann tensor along the orbit on this tetrad frame as in equation~\eref{eq5}. The result may be expressed as a $6\times 6$ matrix $(\mathcal{R}_{AB})$, where $A$ and $B$ are indices that belong to the set $\{ 01,02,03,23,31,12\}$. This matrix has the form
\begin{equation}
\label{eq28} \mathcal{R}=\left[\begin{array}{cc} \mathcal{E} & \mathcal{H} \\ \mathcal{H} & -\mathcal{E}\end{array}\right],
\end{equation}
where $\mathcal{E}$ and $\mathcal{H}$ are $3\times 3$ symmetric and traceless matrices containing respectively the electric and magnetic components of the Riemann tensor. It turns out that in the case under consideration
\begin{equation}
\label{eq29} \mathcal{E}=\kappa \left[ \begin{array}{ccc} k_1 & 0 & k'\\ 0 & k_2 & 0\\ k' & 0 & k_3\end{array}\right],\qquad \mathcal{H} =\kappa \left[ \begin{array}{ccc} 0 & h & 0\\ h & 0 & h' \\ 0 & h' & 0\end{array}\right] ,
\end{equation}
where 
\begin{equation}
\label{eq30} \kappa =\omega^2_K=\frac{M}{r^3}.
\end{equation}
The proportionality $\mathcal{R}\propto \kappa$ is expected on the basis of the correspondence with Newtonian tides. In \eref{eq29}, $k_2=-(k_1+k_3)$ is constant and is given by $k_2=3\gamma^2-2$, while 
\begin{eqnarray}
\label{eq31}k_1=1-3\gamma^2\cos ^2\eta , \quad k_3=1-3\gamma^2\sin ^2\eta, \quad k'=-3\gamma^2\sin \eta \cos \eta,\\
\label{eq32} h=-3\gamma^2\beta \cos \eta ,\quad h'=-3\gamma^2\beta \sin \eta .
\end{eqnarray}
Here $(\beta ,\gamma)$ is a new Lorentz pair, i.e. $\gamma =1/\sqrt{1-\beta ^2}$, given by
\begin{equation}
\label{eq33} \beta =\frac{r^2\omega_K -a}{\sqrt{\Delta}} ,\qquad \gamma =\frac{\sqrt{\Delta}}{rN},
\end{equation}
which reduces to $(\tilde{\beta},\tilde{\gamma})$ in the Schwarzschild limit $(a=0)$; moreover, $\gamma$ diverges at the null orbits ($N=0)$. 

Let us note that for the orbits under consideration and 
$a\le M$, $\beta \in [ -1/2, 1/2]$; that is, $\beta$ is positive (negative) for a prograde (retrograde) orbit and  far from the source
($r\gg 2 M$), $| \beta | \sim \sqrt{M/r} $,  so that $\beta\to 0$ as $r\to\infty$, then $|\beta|$ monotonically increases from zero to $1/2$ at the last stable circular orbits. Furthermore, for $a>M$, $\beta$ is always negative for retrograde orbits, but it is not always positive for prograde orbits; in fact, $\beta$ vanishes for a prograde orbit with radius $r=a^2/M$.

These curvature components can now be used in equations \eref{eq3} and \eref{eq6} to study the motion of the probe relative to $m$. It should be mentioned that our results for the tidal tensor $\mathcal{E}$ are consistent with the work of Marck~\cite{8}.

\section{Jacobi Equation\label{s3}}

Let us now turn to the solution of the Jacobi equation~\eref{eq6}. For simplicity, we use instead of the Fermi coordinates $X^\alpha =(T,\mathbf{X})$, the \textit{dimensionless} Fermi coordinates $(\eta ,\mathbf{x})$,
\begin{equation}
\label{eq34} \omega_K X^\alpha :=(\eta ,\mathbf{x}),
\end{equation}
where $\mathbf{x}=(x,y,z)$. The Jacobi equation then takes the form

\begin{eqnarray}
\label{eq35} \frac{d^2}{d\eta^2} \left[ \begin{array}{c} x \\ z\end{array}\right] =S \left[ \begin{array}{c} x\\ z\end{array}\right], \qquad S=\left[ \begin{array}{cc} 3\gamma^2 \cos ^2\eta-1 & 3\gamma^2 \sin \eta \cos \eta \\ 3\gamma^2 \sin \eta \cos \eta & 3\gamma^2 \sin ^2\eta -1 \end{array}\right],\\
\label{eq36} \frac{d^2y}{d\eta^2} +\sigma^2 y=0,\qquad \sigma =\sqrt{3\gamma^2-2}.
\end{eqnarray}
Thus normal to the equatorial plane, we have a simple harmonic motion of frequency $|\omega_K|\sigma$, where $\sigma$ ranges from 1 to $\sqrt{2}$ for $a\le M$; that is, it increases monotonically
from unity at $r = \infty$ to $\sqrt{2}$ at the last stable circular orbits. The equations of relative motion in the equatorial plane can be put into an autonomous form by a rotation to the radial and tangential directions (cf. \Fref{fig:1}). That is, let
\begin{equation}
\label{eq37} \left[ \begin{array}{c} \xi \\ \zeta \end{array}\right] =R\left[ \begin{array}{c} x \\ z\end{array}\right] ,\qquad R=\left[ \begin{array}{cc} \cos \eta & \sin \eta \\ -\sin \eta & \cos \eta \end{array} \right].
\end{equation}
Then, equation~\eref{eq35} reduces to the autonomous system
\begin{equation}
\label{eq38} \ddot{\xi} -2\dot{\zeta} -3\gamma^2 \xi =0,\qquad \ddot{\zeta} +2\dot{\xi }=0,
\end{equation}
where an overdot denotes differentiation with respect to $\eta$. Assuming that at $\eta=0$, $\mathbf{ x}=0$ and the initial velocity of the probe $(V_0\ll 1)$ has components
\begin{equation}
\label{eq39} \mathbf{V_0} =V_0 (\sin \vartheta \cos \varphi ,\cos \vartheta ,\sin \vartheta \sin \varphi )
\end{equation}
in the local $(X,Y,Z)$ system, we find
\begin{eqnarray}
\label{eq40} \xi = \frac{V_0 \sin \vartheta}{\lambda^2} [2 (1-\cos \lambda \eta )\sin \varphi+\lambda  \sin \lambda \eta \cos \varphi],\\
\label{eq41} y=\frac{1}{\sigma }V_0 \cos \vartheta \sin \sigma \eta ,\\
\label{eq42} \zeta =- \frac{V_0\sin \vartheta}{\lambda^2} [ ( 3\gamma^2\eta -\frac{4}{\lambda} \sin \lambda \eta )\sin \varphi +2 (1-\cos \lambda \eta )\cos \varphi],
\end{eqnarray}
where $\lambda=\sqrt{4-3\gamma^2}$. We note that $\lambda \geq 0$, since
\begin{equation}
\label{eq43} \lambda^2 =\frac{1}{N^2} \big( 1-6\frac{M}{r}+8a\omega_K -3\frac{a^2}{r^2}\big),
\end{equation}
so that for $a\le M$,  $\lambda$ ranges from zero to unity; that is, it decreases monotonically from unity at $r = \infty$ to zero at the last stable circular orbits. If the reference trajectory is one of the last stable circular orbits, then $\gamma^2=4/3$, $y$ is given by \eref{eq41} with $\sigma=\sqrt{2}$ and 
\begin{eqnarray}
\label{eq44} \xi = V_0 \sin \vartheta (\eta\cos \varphi  + \eta^2 \sin \varphi),\\
\label{eq45} \zeta =V_0\sin \vartheta \big[ \big ( \eta -\frac{2}{3} \eta^3\big) \sin \varphi - \eta^2 \cos \varphi\big].
\end{eqnarray}
Recalling the restrictions $|\mathbf{X}|\ll \mathcal{R}_a$ and $|\mathbf{V}|\ll 1$, which in this case translate to $|\mathbf{x}|\ll 1$ and $|\dot{\mathbf x}|\ll 1$, we find that the validity of equations \eref{eq42}, \eref{eq44} and \eref{eq45} is limited in time due to the presence of secular terms.

It is interesting to note that terms that appear in the solution of the Jacobi equation in equations~\eref{eq40} and~\eref{eq42} with proper frequency $\lambda |\omega_K|$ have frequency $\lambda |\omega_K| N/(1+a\omega_K)$ with respect to the coordinate time $t$; to lowest order in $M/r\ll 1$ and with $a = 0$, the deviation of this frequency from the Keplerian frequency corresponds to the Einstein precession frequency for an orbit of vanishing eccentricity, while the first-order correction in $a/r\ll 1$ corresponds to the de Sitter-Lense-Thirring precession frequency~\cite{new14,new15}.

Finally, a remark is in order here regarding the boundary condition that $\mathbf{X}=0$ at $\tau=0$, which corresponds to our assumption that the probe is initially launched from $m$. In fact, it is only necessary that the probe be near the reference particle $m$, so that  $|\mathbf{X}|$ must be initially very small compared to $\mathcal{R}_a$; for the sake of simplicity, we choose $|\mathbf{X}(0)|=0$  throughout this paper. This means that in terms of $(\xi, y,\zeta)$, we always assume that
\begin{eqnarray}\label{eq:50}
(\xi, y,\zeta)\big |_{\eta=0}=0, \qquad (\dot\xi, \dot y,\dot\zeta)\big |_{\eta=0}=\mathbf{V}_0.
\end{eqnarray}

\section{Generalized Jacobi Equation\label{s4}}

The reduction of the Jacobi equation to the autonomous system \eref{eq38} is basically due to the fact that $RSR^{-1}=D$, where $D=\mbox{ diag } (3\gamma^2-1,-1)$ is a constant matrix. System \eref{eq35} then takes the form
\begin{equation}
\label{eq46} \left[ \begin{array}{ccc} \ddot{\xi} & -2\dot{\zeta} & -\xi\\ \ddot{\zeta} & +2\dot{\xi} & -\zeta \end{array} \right] =D\left[ \begin{array}{c} \xi\\ \zeta \end{array}\right] ,
\end{equation}
which is equivalent to system \eref{eq38}. The transformation from  $(x,z)$ to $(\xi ,\zeta)$ involves a rotation from a local inertial frame to rotating axes; therefore, in equation~\eref{eq46} we note the presence of Coriolis and centripetal terms on the left-hand side of this system. Moreover, this transformation is related to Hill's contributions to the classical three-body problem \cite{13,14}; indeed, system \eref{eq38} in the Newtonian limit $\gamma \to 1$ is equivalent to a limiting form of the Hill system discussed in ref. \cite{13}, namely, system \eref{eq36} of ref.~\cite{13} with $k=e_1=0$.

It can be shown, after much algebra, that under the same rotation as in equation \eref{eq37}, the nonlinear generalized Jacobi equation \eref{eq3} can be rendered autonomous as well. This remarkable fact is due to the special symmetries of the Kerr spacetime~\cite{16}. Introducing $\Delta_\xi =\dot{\xi} -\zeta $ and 
$\Delta_\zeta =\dot{\zeta} +\xi$, where
\begin{eqnarray}
\label{eq47} \left[\begin{array}{c} \Delta_\xi \\ \Delta_\zeta \end{array}\right] =R\left[ \begin{array}{c} \dot{x}\\ \dot{z}\end{array}\right],
\end{eqnarray}
we find that $x\dot x+z\dot z=\xi \Delta_\xi+\zeta\Delta_\zeta$, $x\dot z-z\dot x=\xi\Delta_\zeta-\zeta\Delta_\xi$, and
\begin{eqnarray}
\fl \ddot{\xi}-2\dot{\zeta} -3\gamma^2 \xi = 
2\Delta_\xi [\zeta \Delta_\zeta -(\sigma^2+1)\xi \Delta_\xi ] +6\gamma^2\beta \xi \Delta_\zeta\nonumber \\
- 2 \big( \gamma^2\beta \Delta^2_\xi -\frac{1}{3} \sigma^2 \Delta_\zeta \big) 
(\xi \Delta_\zeta -\zeta \Delta_\xi )
 +2 y\dot{y} \Delta_\xi \big( \gamma^2\beta \Delta_\zeta +\sigma^2-\frac{1}{3}\big )\nonumber\\ 
 -2\dot{y}^2 \big (\gamma^2\beta \zeta \Delta_\xi
-\frac{1}{3} \xi \big ),\label{eq48}\\
\fl \ddot{y} +\sigma^2 (1-2\dot{y}^2)y = \frac{2}{3} y[\Delta^2 _\xi -(\sigma^2+1) \Delta^2_\zeta -9\gamma^2 \beta \Delta_\zeta ]
 -2\dot{y} \big[ \gamma^2\beta \Delta_\xi (\xi \Delta_\zeta
-\zeta \Delta_\xi )\nonumber\\
+\big( 3\gamma^2-\frac{2}{3}\big) \xi \Delta_\xi -\big(\gamma^2+\frac{2}{3}\big) \zeta \Delta_\zeta \big]
+2\gamma^2\beta \dot{y}^2(y\Delta _\zeta -\zeta \dot{y}),\label{eq49}\\
\fl\ddot{\zeta}+2\dot{\xi}  = 2\Delta_\zeta [\zeta \Delta_\zeta -(\sigma^2+1)\xi \Delta_\xi ] -6\gamma^2\beta \xi \Delta_\xi
 -2 \Delta_\xi \big( \gamma^2\beta \Delta _\zeta +\frac{1}{3} \sigma^2\big) (\xi \Delta_\zeta -\zeta \Delta_\xi )\nonumber\\
 +\frac{2}{3} (\sigma^2+1)\dot y (y\Delta_\zeta -\zeta \dot{y})+2y\dot{y} [\gamma^2\beta (\Delta^2_\zeta +3)
+\sigma^2\Delta_\zeta ]
-2\gamma^2\beta \dot{y}^2\zeta \Delta_\zeta .\label{eq50}
\end{eqnarray}
Starting with the exact solution of the Jacobi equation as the unperturbed solution, it is straightforward to develop a solution of equations \eref{eq48}--\eref{eq50} via the standard perturbation expansion in terms of $V_0\ll 1$. However, the results of such a perturbation scheme, based on the
small parameter $V_0$, should be used only in conjunction with the higher-order
tidal terms---that is, terms that have been neglected in the expansions of
the gravitational potentials in equations~\eref{eq12a}--\eref{eq14a} and would result in terms
in the equation of motion~\eref{eq1} that go beyond the linear order in $|\mathbf{X}|$.

It follows from equations~\eref{eq:10}, \eref{eq24a}--\eref{eq27}, \eref{eq34} and~\eref{eq37}  that the trajectory of the probe in Boyer-Lindquist coordinates, $x^\mu_p=(t_p, r_p, \theta_p, \phi_p)$, is given by
\begin{eqnarray}
\label{eqn:57} t_p(\tau)=\frac{1}{N}(1+a\omega_K)\tau+\frac{L}{\sqrt{\Delta}\,\omega_K }\,\zeta,\\
\label{eqn:58} r_p(\tau)=r+ \frac{\sqrt{\Delta}}{r \omega_K}\,\xi,\\
\label{eqn:59} \theta_p(\tau)=\frac{\pi}{2}+\frac{1}{r \omega_K}\, y,\\
\label{eqn:60} \phi_p(\tau)=\frac{1}{N}\omega_K\tau +\frac{E}{\sqrt{\Delta}\,\omega_K}\,\zeta.
\end{eqnarray}
Using $\xi(\eta)$, $y(\eta)$ and $\zeta(\eta)$ from the (generalized) Jacobi equation and replacing $\eta=\omega_K T$ by  $\omega_K \tau$, we can determine the path of the probe in the standard Kerr coordinate system. This is due to the fact that the generalized Jacobi equation is linear in the distance $|\mathbf{ X}|$ away from the reference geodesic; therefore, $X^i\lambda^\mu_{\hspace{\myspace}(i)}$ may be viewed as a generalized Jacobi field that is defined along the reference geodesic. It turns out that for small-amplitude perturbations $(V_0\ll1$), the radial ($\xi$) and vertical ($y$) motions in general contain the basic proper epicyclic frequencies $\lambda|\omega_K|$ and $\sigma|\omega_K|$, respectively.

The autonomous system of equations \eref{eq48}--\eref{eq50} naturally splits into equations for the vertical and equatorial motions. That is, if the probe is launched vertically in the $y$ direction, then its
motion will be confined to the $y$ direction according to the generalized
Jacobi equation. Similarly, if the probe is launched in the orbital $( \xi,
\zeta )$ plane, then the probe will remain in this plane throughout its
motion.

\subsection*{Vertical Motion}

If the probe is launched in the purely vertical direction relative to the reference orbit, then $\xi (\eta )=\zeta (\eta )=0$ for all $\eta$ and hence the equations of motion reduce to 
\begin{equation}
\label{eq51} \ddot{y}+\sigma^2 (1-2\dot{y}^2)y=0.
\end{equation}
The vertical motion is simply uniform for $\dot{y}=\pm 1/\sqrt{2}$; that is, equation~\eref{eq51} exhibits the critical speed $V_c=1/\sqrt{2}$ beyond which the character of the motion is opposite to the low-speed limit that agrees with Newtonian expectations. If the probe is launched with $V_0<V_c$, then it will begin to  decelerate in agreement with Newtonian gravity. However, for $V_0>V_c$, the probe will accelerate, as
treated in detail in~\cite{2,4}. In fact, the vertical motion can be described exactly~\cite{2,newbm,4}, since equation~\eref{eq51} implies that
\begin{equation}\label{eq52}
\dot y^2=V_c^2+(V_0^2-V_c^2) e^{2\sigma^2y^2},
\end{equation}
whose solution can be determined by quadrature.
This solution represents timelike geodesic motion if the following condition
is satisfied
\begin{equation}\label{eq53}
\frac{1}{\Gamma^2}=1-\dot y^2+\sigma^2 y^2>0.
\end{equation}
For $V_0^2<V_c^2$, condition~\eref{eq53} is always satisfied and it follows from~\eref{eq52}
that the motion of $y$ is periodic and confined to the interval
$[-y_{\mbox{\footnotesize max}}, y_{\mbox{\footnotesize max}}]$ , where
\begin{equation}\label{eq54}
y_{\mbox{\footnotesize max}}=\frac{1}{\sigma}\sqrt{
\frac{1}{2}\ln\Big(\frac{V_c^2}{V_c^2-V_0^2}\Big)}.
\end{equation}
On the other hand, for $V_0^2>V_c^2$  the particle is accelerated to (almost) the
local speed of light. In this way, the particle gains enormous tidal energy.
According to condition~\eref{eq53}, the range of $y$ over which this happens
monotonically shrinks to zero as $V_0$ approaches unity. The question of the
admissibility of Fermi coordinates in such a case is a difficult one; an
unsuccessful attempt to answer such a question is contained in our recent
work~\cite{3}. It is essential---for astrophysical applications---to connect
these local results to what distant observers would measure~\cite{4}. Indeed, the
gravitational tidal energy of particles within the Fermi coordinate system
can be transferred to the outside world (i.e. beyond the Fermi system)
through collisions with other particles or emission of radiation. 

\begin{figure}
\centerline{\psfig{file=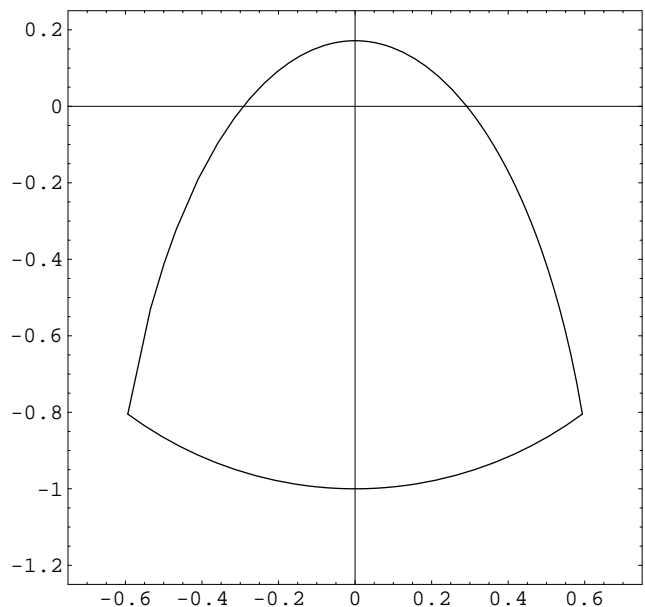, width=20pc}}
\caption{Plot of the boundary of the domain---in the $( \hat{x}, \hat{y})$ plane---in which the vertical tidal acceleration of the probe is positive for $V_0 > 1/\sqrt{2}$. The probe is launched in the direction $( \vartheta, \varphi )$ with respect to the local frame of the reference observer. We define $\hat{x} := \rho \cos \varphi$, $\hat{y}:= \rho \sin \varphi$, where $\rho = \sin \vartheta$. For this plot, $V_0 = 0.8$ and $\beta = 0.45$. The upper part of the boundary curve is given by equation~\eref{eqn:69}. The lower part of the boundary curve is an arc of the circle of radius $\rho = 1$ on the  interval $\pi + \alpha \le \varphi \le 2 \pi - \alpha$, where $\alpha \approx 0.94$  radian corresponding to about $53^\circ$.\label{fig:1.5}}
\end{figure}

In our previous work involving tidal dynamics about a reference escape trajectory along the axis of rotation of the central source~\cite{4}, we found that for relative speeds above the critical speed $V_c$,  tidal \emph{deceleration} occurs in a cone of angle $\Theta$ measured from the rotation axis, where $\Theta$ is given by  $\tan \Theta = 1 / V_c$  and corresponds to an angle of about $55^\circ$, while tidal \emph{acceleration} occurs outside this cone, i.e. within the complement of the
critical-velocity cone involving latitudes from $-\Theta'$ to $\Theta'$, where $\Theta' = \pi/2 - \Theta $  and corresponds to an angle of about $35^\circ$.  Based on these previous results, we expect that in the present situation the phenomenon of tidal acceleration for speeds above the critical speed may not be restricted to the vertical axis alone. To investigate
this issue in the case under consideration, we have integrated the full
system of equations~\eref{eq48}--\eref{eq50} with initial conditions~\eref{eq:50}. Our results indicate that vertical acceleration---either parallel or antiparallel to the Kerr rotation axis---occurs when the probe is launched
with $V_0 > 1/\sqrt {2}$ in a direction $( \vartheta, \varphi )$, where for a given $\varphi\in [0, 2\pi)$, $\vartheta$ is restricted to a certain interval, 
namely, $\vartheta\in [0, \vartheta_0)$. To clarify the situation analytically, let us first note that if $( \xi, y, \zeta )$ is a
solution of equations~\eref{eq48}--\eref{eq50}, then so is $( \xi, - y, \zeta )$. In
particular, equation~\eref{eq51} is clearly invariant under the parity
transformation $y\mapsto - y$; therefore, vertical motion is expected to occur
symmetrically with respect to the equatorial plane. For this reason, we
limit our considerations to $y > 0$, which corresponds to motion antiparallel
to the Kerr axis in our convention (see Figure~\ref{fig:1}). Furthermore, let us note that with
initial conditions~\eref{eq:50}, vertical acceleration vanishes initially, i.e.
$\ddot y (\eta) = 0$ at $\eta = 0$; therefore, it proves useful to compute
$\tdot{y}(0)$ using equation~\eref{eq49}. It turns out that
\begin{eqnarray} \label{eqn:65} 
\tdot{y} (0) = 2 V_0  W ( \vartheta, \varphi )\cos \vartheta,    \end{eqnarray}
where $W$ is a quadratic function of $\rho = \sin \vartheta$ given by
\begin{eqnarray}  \label{eqn:66} 
W = C_0 - V_0 C_1 \rho - V_0^2 C_2 \rho^2.       
\end{eqnarray} 
Here $C_0$, $C_1$ and $ C_2$ are defined by
\begin{eqnarray} \label{eqn:67}  C_0 = \sigma^2 (V_0^2 - \frac{1}{2} ), \qquad 
C_1 = 3 \gamma^2 \beta \sin \varphi, \\
\label{eqn:68}  C_2 = ( 2 \sigma^2 + 1) - ( \sigma^2 + 2 ) \sin^2 \varphi.   
\end{eqnarray}
For $\vartheta = 0$, $\tdot{y}(0)$ is positive for $V_0 > 1/\sqrt {2}$; in
fact, this is the case for an extended domain about the Kerr axis. The
boundary of this domain is characterized by $W = 0$, i.e. $\rho(\varphi)=\sin \vartheta_0$ given by
\begin{eqnarray}\label{eqn:69}   
\rho ( \varphi ) = \frac{1}{2 V_0 C_2}( \sqrt{C_1^2 + 4 C_0 C_2} - C_1), 
\end{eqnarray} 
and  $\rho = 1$ (i.e. $\vartheta_0=\pi/2$), where $\tdot{y}(0)$ vanishes, as illustrated in Figure~\ref{fig:1.5},  where $V_0 =
0.8$ and $\beta = 0.45$. Let us recall that for a prograde reference orbit of
radius $r$ about a Kerr black hole, $\beta$ increases monotonically from zero at $r= \infty$ to $0.5$ at the last stable circular orbit. For a fixed $\beta \in  (0,0.5]$ and $V_0>V_c$, the tidal-acceleration domain expands as $V_0\to 1$. On the other hand, for a fixed $V_0\in (V_c,1)$,
the acceleration domain in Figure~\ref{fig:1.5} remains practically the same as $\beta\to 1/2$. For $\beta\to 0$, however, the shape of the domain changes significantly. In fact, for $\beta=0$, it follows from equation~\eref{eqn:69} that $\rho=\rho_0/|\cos\varphi|$, where $\rho_0$ defined by  $\sqrt{3}\,\rho_0=\sqrt{1-V_c^2/V_0^2}\,$ is a
positive constant such that $\rho_0 \in  (0, 1/\sqrt{6})$. 
Therefore, the acceleration domain in this case is within the intersection of the vertical strip defined by $- \rho_0 \le\hat{x}\le 
\rho_0 $, where $\hat{x} = \rho \cos \varphi$ and $\hat{y} = \rho \sin \varphi$, and the circle of unit radius centered at the origin of $(\hat{x},\hat{y})$ coordinates.

      Though we have considered a simple circular equatorial orbit as the
reference trajectory in this paper, it is not difficult to imagine that the
tidal acceleration process described above---within the context of the
physics of an accretion disk about the source---could possibly contribute
to the formation of astrophysical jets. Moreover, tidal acceleration occurs symmetrically with respect to the
equatorial plane of the Kerr source. This feature would be consistent with the observed occurrence of
double jets, namely, a pair of relativistic outflows in opposite directions
along the rotation axis of the central source.

\subsection*{Equatorial Motion}
\begin{figure}
\centerline{\psfig{file=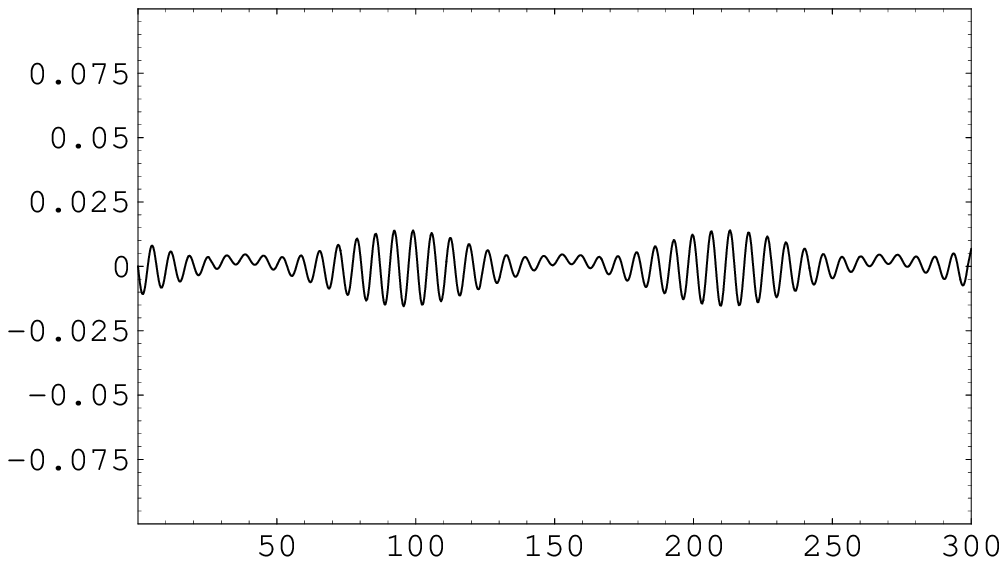, width=15pc}\psfig{file=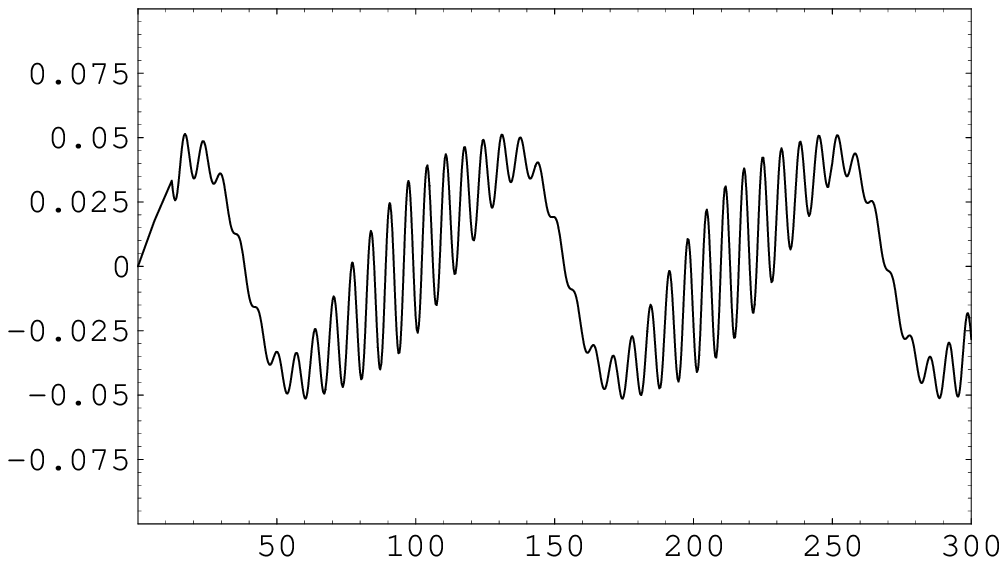, width=15pc}}
\centerline{\psfig{file=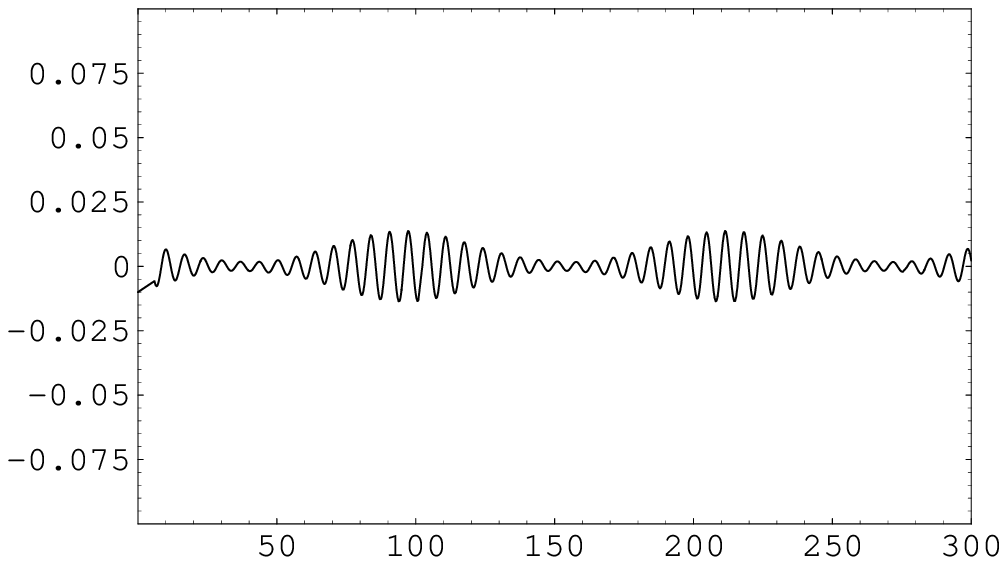, width=15pc}\psfig{file=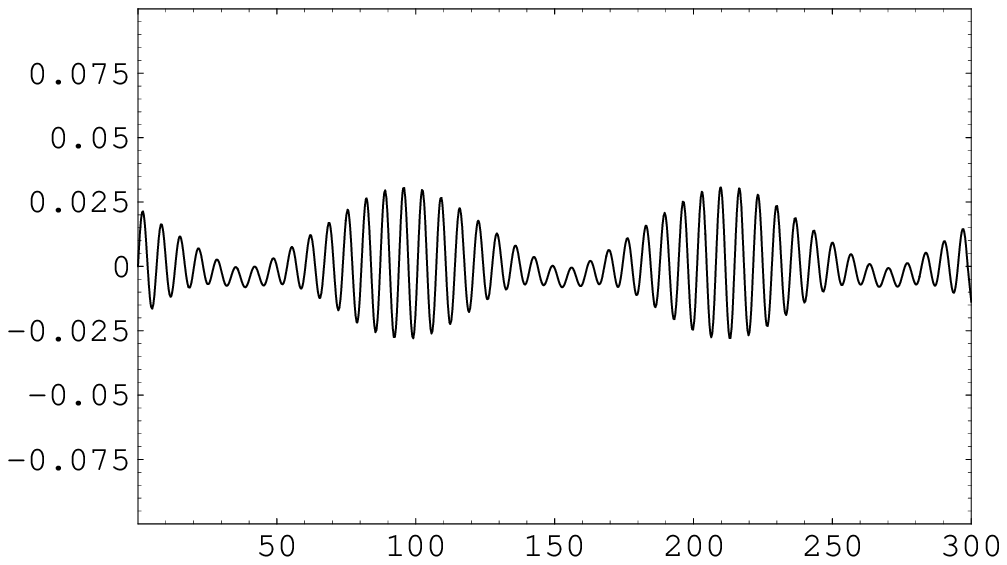, width=15pc}}
\caption{Plots of $(\xi,\zeta)$ versus $\eta$, top panels, and $(\dot\xi,\dot\zeta)$ versus $\eta$, bottom panels, for the generalized Jacobi equation with $\beta=1/5$. The initial conditions are such that the motion is confined to the equatorial plane with a radially inward initial velocity such that $\vartheta=\pi/2$, $\varphi=\pi$ and $V_0=0.01$. Note that the fast oscillation in the above panels has essentially the same frequency as in the corresponding solutions of the Jacobi equation, $(\xi_J,\zeta_J)$, given by $\xi_J=-(V_0/\lambda)\sin\lambda \eta$ and  $\zeta_J=(2 V_0/\lambda^2)(1-\cos\lambda \eta)$.  \label{fig:2}}
\end{figure}
If $y(\eta )=0$ for all $\eta$, then the motion is confined to the $(\xi ,\zeta )$ plane and represents a timelike geodesic if 
\begin{eqnarray}\label{eq:59}
\fl \frac{1}{\Gamma^2}=1-\Delta_\xi^2-\Delta_\zeta^2-(\sigma^2+1)\xi^2+\zeta^2-4\gamma^2\beta\xi( \xi \Delta_\zeta-\zeta\Delta_\xi )-\frac{1}{3}\sigma^2( \xi \Delta_\zeta-\zeta\Delta_\xi )^2>0.
\end{eqnarray}
A solution of the generalized Jacobi equation (GJE) in the $(\xi,\zeta)$ plane is characterized by the parameters $\beta$, $V_0$  and $\varphi$. For  $V_0$ extremely small compared to unity, the solution of the GJE is given essentially by the corresponding solution of the Jacobi equation as in the previous section. In general, as  $V_0$ slowly increases toward unity, the influence of the nonlinear terms in the GJE cannot be ignored. In fact, since $\zeta$ in the Jacobi case contains a secular term, namely,  $-3(\gamma/\lambda)^2(V_0\sin\vartheta\sin\varphi)\eta$ with $\vartheta=\pi/2$, the solution of the GJE in general grows rapidly and leaves the admissible region of the Fermi coordinate system. This difficulty can be avoided, however, in the case of the purely radial variations with $\varphi=0$  or $\pi$; in either case, the Jacobi equation has periodic solutions corresponding to the stability of the circular reference orbit under small radial perturbations. The solution of the GJE may then contain quasi-periodic oscillations exhibiting a complex beat phenomenon involving several frequencies as demonstrated
in Figure~\ref{fig:2}. However, such oscillations appear to occur only for $V_0 \ll 1$, in
which case equations~\eref{eq48}--\eref{eq50} are generally not adequate physically, since
higher-order tidal terms of comparable magnitude have been neglected in our
analysis. Even if a quasi-periodic oscillation of the type illustrated in
Figure~\ref{fig:2} survives the inclusion of relevant higher-order tidal terms, the
amplitude of the effect may not be large enough to be of any physical
significance in connection with the observed  quasi-periodic oscillations (QPOs)~\cite{12}. 

\section{Discussion\label{s5}}  
We have studied the generalized Jacobi equation for a circular reference geodesic orbit in the equatorial plane of the exterior Kerr spacetime. This equation has been reduced to an autonomous system; the corresponding tidal dynamics can then be naturally divided into vertical and equatorial motions relative to the circular orbit. In connection with vertical motion,  we have clarified the role of the critical speed $V_c=1/\sqrt{2}$ in the character of the solutions of the generalized Jacobi equation.

The  general approach developed in this work may be of interest in connection with relativistic physics inside a space station, the tracking of artificial Earth satellites as well as satellite-to-satellite Doppler tracking. Moreover, in connection with the relativistic astrophysics of accretion disks around (rotating) astronomical sources, our results may be relevant for the gravitational aspects of the complex plasma physics  that would be involved in the formation of high-energy jets~\cite{new20}. That is, the vertical acceleration phenomenon explored in this work may be related to how relativistic jets get started above and below an accretion disk around a Kerr black hole. Once the flows are magnetically confined to regions near the axis of
rotation of the central source, the results of previous investigations~\cite{ 4,5,11} involving tidal dynamics of ultrarelativistic particles would have
to be taken into account for the parallel and antiparallel flows along the
Kerr rotation axis.

\end{document}